# Probing critical point energies of transition metal dichalcogenides: surprising indirect gap of single layer WSe$_2$

*Chendong Zhang[1], Yuxuan Chen[1], Amber Johnson[1], Ming-Yang Li[2], Lain-Jong Li[3], Patrick C. Mende[4], Randall M. Feenstra[4] and Chih-Kang Shih[1*]*

[1]*Department of Physics, University of Texas at Austin, Austin, TX 78712, USA*
[2] *Institute of Atomic and Molecular Sciences, Academia Sinica, No. 1, Roosevelt Rd., Sec. 4, Taipei 10617, Taiwan*
[3]*Physical Sciences and Engineering Division, King Abdullah University of Science and Technology, Thuwal, 23955-6900, Kingdom of Saudi Arabia.*
[4]*Department of Physics, Carnegie Mellon University, Pittsburgh, PA 15213, USA*

*\*Corresponding author E-mail: shih@physics.utexas.edu*

**Understanding quasiparticle band structures of transition metal dichalcogenides (TMDs) is critical for technological advances of these materials for atomic layer electronics and photonics. Although theoretical calculations to date have shown qualitatively similar features, there exist subtle differences which can lead to important consequences in the device characteristics. For example, most calculations have shown that all single layer (SL) TMDs have direct band gaps, while some have shown that SL-WSe$_2$ have an indirect gap. Moreover, there are large variations in the reported quasiparticle gaps, corresponding to large variations in exciton binding energies. By using a comprehensive form of scanning tunneling spectroscopy, we have revealed detailed quasiparticle electronic structures in TMDs, including the quasi-particle gaps, critical point energy locations and their origins in the Brillouin Zones (BZs). We show that SL-WSe$_2$ actually has an indirect quasi-particle gap with the conduction band minimum located at the Q point (instead of K), albeit the two states are nearly degenerate. Its implications on optical properties are discussed. We have further observed rich quasi-particle electronic structures of TMDs as a function of atomic structures and spin-orbital couplings.**

Experimental determinations of the electronic band structures in TMDs are quite non-trivial. Optical spectroscopies[1-5] are unsuitable to measure the quasi-particle band structures due to the existence of large exciton binding energies. Using angle resolved photoemission (ARPES), it is difficult to probe the conduction band structures[6,7]. In principle, scanning tunneling spectroscopy (STS) would be an ideal probe to determine both the valence and conduction band structures. However, the reported results have been controversial thus far, even for the determination of the quasi-particle band gaps[8-10]. As we will show, this is due primarily to the intriguing influence of the lateral momentum in the tunneling process, making certain critical points difficult to access in the conventional scanning tunneling spectroscopy acquired at a constant tip-to-sample-distance (Z). By using a comprehensive approach combining the constant $Z$ and variable $Z$ spectroscopies, as well as state-resolved tunneling decay constant measurements, we have shown that detailed electronic structures, including quasi-particle gaps, critical point energy locations and their origins in the BZs in TMDs can be revealed.

The TMD samples are grown using chemical vapor deposition (CVD) for WSe$_2$[11] or molecular beam epitaxy (MBE) for MoSe$_2$ on highly-oriented-pyrolytic-graphite (HOPG) substrates. In addition, MoSe$_2$ has also been grown on epitaxial bi-layer graphene to investigate the environmental influences on the quasi-particle band structures. Figures 1a and b show scanning tunneling microscopy (STM) images of MoSe$_2$ and WSe$_2$, respectively. Due to a nearly 4:3 lattice match with the graphite, the TMD samples also show Moiré patterns with a periodicity of ~ 1nm. An example is shown as an inset in Fig. 1b, similar to those reported earlier[9]. In Fig. 1c we show a generic electronic structure for SL-TMD materials. We first discuss the result of MoSe$_2$ due to the availability of experimentally determined *E vs. k* dispersion in the valence band which can be used to cross-check with our results.

Figure 2a shows a typical constant Z (tip-to-sample distance) differential conductivity (*dI/dV*) spectrum (displayed in log-scale) acquired on SL-MoSe$_2$. The peak at around -1.9 V is the energy location of the Γ point. However, the electronic states near the valence band maximum (VBM) at the K point could not be observed. The failure to detect such states is due to their much larger parallel momentum ($k_\parallel$) value in the BZ, thus resulting in a larger tunneling decay constant. Following the original theory of Tersoff[12,13], for an electronic state with a parallel momentum of $k_\parallel$, the effective tunneling decay constant is $= \sqrt{\frac{2m\Phi_b + k_\parallel^2}{\hbar^2}}$, where $\Phi_b$ is the energy barrier for tunneling. A typical barrier height is between 2 and 4 eV (depending on the bias), resulting in a typical decay constant of $\kappa = 0.7$~$1.0$ Å$^{-1}$ for $k_\parallel = 0$. In SL-MoSe$_2$, the VBM is located at K with a $k_\parallel$ of 1.28 Å$^{-1}$, yielding a significantly larger value of $\kappa = 1.4$ ~$1.6$ Å$^{-1}$.

Such a large difference in the decay constant is responsible for the difficulty in detecting the states near the VBM. The lack of the sensitivity in the constant-$Z$ STS can be overcome by acquiring spectra at variable $Z$ as described before[14]. Here we adopt a form of variable $Z$ spectroscopy by performing STS at constant current. In this mode, as the sample bias is scanned across different thresholds, $Z$ (the dependent variable) will respond automatically in order to keep the current constant. The differential conductivity $(\partial I/\partial V)_I$ is measured by using a lock-in amplifier (Fig. 2b). In the meantime, the acquired Z-value is used to deduce $(\partial Z/\partial V)_I$ which can be used to identify individual thresholds (Fig. 2c).

For the valence band (left panel), the state at the Γ point also appears as a prominent peak in the $(\partial I/\partial V)_I$ spectrum. Moreover, spectroscopic features above Γ are observed due to the significant enhancement of the sensitivity. A shoulder at ~ 0.15 eV above Γ is identified as the state at $K_2$ (the lower energy state at the critical point K below the VBM as marked in Fig. 1c). Above $K_2$ are states near the global VBM at the K point (labeled as $K_V$). The energy separation between $K_2$ and $K_V$ is due to the spin-orbital coupling (labeled as $\Delta_{SO}$). The ability to detect these quickly decaying states is due to the fact that the tip-to-sample-distance is brought closer by the feedback circuit, reflected by the accompanying $(\partial Z/\partial V)_I$ spectrum in Fig. 2c. At the Γ point, one can observe a sharp dip in the $(\partial Z/\partial V)_I$ spectrum, corresponding to a sudden reduction in $Z$. This occurs due to the loss of the available states at the Γ point in the tunneling window, leaving only states with a higher decay constant than that at Γ. A sudden reduction of $Z$ (appearing as a sharp dip in $(\partial Z/\partial V)_I$) occurs in order to maintain the constant current. This underlies the principle behind the sensitivity of $(\partial Z/\partial V)_I$ to the onset of individual thresholds in the density of states (DOS). Another clear $Z$ reduction occurs at around -1.5 V corresponding to the location of the global VBM at the K point. Above this voltage, the tunneling no longer occurs between the MoSe$_2$ electronic states and the tip. Rather, the tunneling is due to the underlying graphite states and MoSe$_2$ becomes part of the tunneling barrier. Similarly the conduction band electronic states (right panel) can be revealed with greater sensitivity than the conventional constant Z spectroscopy. In this case, the individual threshold shows as a peak (instead of a dip), in the $(\partial Z/\partial V)_I$ spectrum, since the threshold is crossed by scanning the sample bias toward smaller positive voltage. In order to gain insight on their origin in the BZ for these different thresholds, we also measure the state-resolved tunneling decay constant. Experimentally, this is determined by measuring the logarithmic derivative of tunneling current w.r.t Z as $\kappa = -\frac{1}{2}\frac{\partial \ln I}{\partial Z}$, as shown in Fig. 2d.

Recalling that the effective tunneling decay constant is $\kappa = \sqrt{\frac{2m\Phi_b + k_\parallel^2}{\hbar^2}}$, one thus expects to see a minimum in $\kappa$ at $\Gamma$ and a maximum in $\kappa$ at K (for MoSe$_2$, $k_\parallel$ = 1.28 Å$^{-1}$). Indeed this qualitative behavior is observed: a sharp dip in $\kappa$ coincides with the assignment of $\Gamma$ and rises to a maximum value near K. We should mention that this effect of $k_\parallel$ on the tunneling decay constant has been first observed in Si(111)-2x1 surfaces[15]. In the current case, due to the electronics states from the underlying graphite, one can scan directly across the VBM and observe a finite transition width from the VBM of MoSe$_2$ to the underlying graphite states.

To capture the quantitative behaviors, we have carried out theoretical simulations using a four band model, including two bands that are degenerate at the $\Gamma$ points and two spin-orbit split bands near K( as shown in the top panel of Fig. 2e). Here we use only parabolic bands and a value of $\Delta_{SO}$ = 0.23 eV for spin-orbital splitting and an energy separation of $\Delta_{K-\Gamma}$ of 0.38 eV from $\Gamma$ to VBM. The bias dependent tunneling barrier height is approximated as $\Phi_b = \Phi_o -$

e|$V_s$|/2 with $\Phi_o$ = 4 eV. In addition, the underlying graphite electronic states are approximated by using a graphene-like, linearly energy-dependent DOS, located at ~ 5 Å below that of TMD (more discussions in the supplementary information). The simulated results for $(\partial I/\partial V)_I$, $(\partial Z/\partial V)_I$, and $\kappa = -\frac{1}{2}\frac{\partial \ln I}{\partial Z}$ as a function of sample bias for filled states are shown in Fig. 2e, exhibiting excellent resemblance to the experimental results. For states below the Γ point, the simulation shows much more gradual variation than the experiment, due primarily to the simplified 4 band approximation which does not capture the actual E vs. k dispersion below Γ. But this approximation should not impact our ability to quantitatively determine the energy locations of the critical points. The simulation also helps us to correctly identify the Γ point occurring at the dip location in the $(\partial Z/\partial V)_I$ and the $K_V$ point occurring at the mid-point of the transition from the TMD to the graphite states (more discussions in supplementary).

Combining the experimental measurements and numerical simulations for MoSe$_2$, the critical points are determined as $\Gamma_V$ = -1.87 ± 0.03 eV, $K_2$ = -1.72 ± 0.05 eV and $K_V$ = -1.48 ± 0.03 eV. Note also, our determination of $\Delta_{\Gamma-K}$ = 0.39 ± 0.04 eV, is in excellent agreement with the recent ARPES measurement of MoSe$_2$[6]. In addition, the experimental value of $\Delta_{SO}$ = 0.24 ± 0.06 eV, also agrees very well with most theoretical calculations[16-18]. Similar measurements of the conduction band for SL-MoSe$_2$ are shown in the right panel of Fig. 2b-d. Three thresholds are observed. The lowest energy threshold at 0.67 ± 0.03 eV can be identified as the conduction band minimum (CBM) occurring at the K-point (highest $\kappa$ value). The threshold (appearing as a dip) at 0.86 ± 0.03 eV is attributed to the $Q_C$ point located half way between Γ and K, while the broad peak near 1.0 ± 0.1 eV is attributed to states between Γ and M (containing several bands). The quasi-particle gap, as determined experimentally, would be 2.15 ± 0.06 eV for MBE grown

SL-MoSe$_2$ on graphite. Since both VBM and CBM occur at the K point, the result confirms a direct band gap for SL-MoSe$_2$. We note that our value is very close to a recent reported value for SL-MoSe$_2$ on bi-layer graphene[9]. However, in reference 9, it was also reported that for MoSe$_2$ on graphite there is a reduction of quasiparticle gap by 0.22 eV, which is different from our result for MoSe$_2$ on graphite. In order to resolve this inconsistency, we have investigated MoSe$_2$ grown on bi-layer graphene and observed a similar value as from our measurements on HOPG for the quasi-particle gap and the relevant critical point energies (see supplementary information). Note that the extreme two-dimensional (2D) nature of SL-TMDs makes them particularly sensitive to the coupling to the substrates. We suggest that the difference between our results and those of reference 9 for the MoSe$_2$ on HOPG is likely due to the details in the growth. In our sample systems, quasi-particle band gaps are the same whether MoSe$_2$ is on bi-layer graphene or on graphite, implying very similar coupling of MoSe$_2$ to the substrates.

The STS of WSe$_2$ on graphite (CVD grown) is also investigated. In the regular constant Z $dI/dV$ spectrum, a prominent peak at the $\Gamma_V$ point is also observed (Fig. 3a). However, the states near the VBM at the K point are now completely absent. On the other hand, in the $(\partial I/\partial V)_I$ spectrum the states near the VBM at the K point can be observed quite clearly. The general behavior of valence band for WSe$_2$ is similar to that for MoSe$_2$ with $\Gamma_V$, $K_2$, and $K_V$ determined to be -1.64 ± 0.03 eV, -1.44 ± 0.05 eV, and -1.00 ± 0.05 eV, respectively. Again, we determine $\Delta_{K-\Gamma}$ = 0.64 ± 0.06 eV and $\Delta_{SO}$ = 0.44 ± 0.06 eV in WSe$_2$, consistent with theoretical calculations[16-18]. The companion simulation results are shown in Fig. S2. The larger $\Delta_{SO}$ in WSe$_2$ than that in MoSe$_2$ is a direct consequence of the heavier TM atom.

The conduction band structures (right panels) show some intriguing differences. Unlike the case for MoSe$_2$ where multiple thresholds are identified near the CBM, for WSe$_2$, only one prominent peak at 1.15 eV is observed near CBM in $(\partial I/\partial V)_l$ although a weak shoulder at about 1.20 eV is also observed. Above this threshold, a broad peak around 1.5 eV can be assigned as the states near the M point. Many theoretical calculations show that there are two nearly degenerate critical points at the CBM: one at K and the other one at Q which is located half way between Γ and K[5,16,17,19,20]. Some show that CBM is at K, however some calculations actually predict that CBM is located at Q[16,17]. The closeness in energy between Q and K is the reason why we only observe a single peak in $(\partial I/\partial V)_l$ and $(\partial Z/\partial V)_l$. But the most interesting aspect is the behavior of $\kappa$, the tunneling decay constant: as one moves from a higher energy state (say 1.5 eV) down toward the CBM, one first observes a rising $\kappa$ (indicative of a large $k_\parallel$), which is then followed by a sharp dip in κ right near the CBM position. This would indicate that the CBM is not at the K point. This is to our knowledge the first direct experimental evidence for a Q–K reversal in SL-WSe$_2$ near the CBM. We have also investigated the systematic evolution of $\Delta_{Q-K}$ in four different SL-TMDs compounds (Fig. S4) and find that WSe$_2$ is the only one with the CBM at the Q point. For WSe$_2$, the CBM position determined using $(\partial Z/\partial V)$ would yield a value of 1.12 ± 0.03 eV at the Q point. We can also estimate that the K-point is located slightly higher, roughly at 1.20 ± 0.05 eV. Thus, based on our experimental results, SL-WSe$_2$ has an indirect quasi-particle gap (from Q to K), of 2.12 ± 0.06 eV. The observation of "indirect gap" in SL-WSe$_2$ is quite surprising as the SL-WSe$_2$ is known for its efficient photoluminescence. We note however, that there is a direct quasi-particle gap at the K-K transition of 2.20 ± 0.10 eV, nearly degenerate with the indirect gap. This nearly degenerate direct gap, plus the fact that the exciton binding energy is quite large (~ 0.5 eV), may be responsible for the efficient optical transition. Another possibility is that this Q-K reversal of CBM is due to the fact that our WSe$_2$ is grown on graphite with a Moiré pattern which imposes additional periodicity, thus modifying the quasi-particle band structure. This interesting scenario warrants further investigations, in particular with the new capability to probe detailed band structure offered by the comprehensive STS.

The direct K-K transition in WSe$_2$ is ~ 50 meV higher than that of MoSe$_2$, although this value contains a larger uncertainty due to the large error in the determination of K$_C$ point from the shoulder. In comparison, on the same sample, photoluminescence measurements at 79 K show an exciton transitions of 1.71 eV for SL-WSe$_2$ which is 0.08 eV higher than the transition of 1.63 eV for SL-MoSe$_2$ on graphite (supplementary information), consistent with the difference in the direct quasiparticle gap (K-K transition). Based on the measured quasi-particle gaps and the exciton transition energies, we find a similar exciton binding energy of ~ 0.5 eV for MoSe$_2$ and WSe$_2$.

We further investigate the effect of interlayer coupling in double layer (DL) TMDs using STS (Fig. 4). The energy splittings, $\Delta_{\Gamma-\Gamma}$, at the $\Gamma$ point in VB, are measured to be 0.72 ± 0.1 eV for MoSe2 (similar to ref. 21) and 0.76 ± 0.1 eV for WSe$_2$. For the conduction band, the $(\partial Z/\partial V)_I$ and $\kappa$ measurements (Fig. 4b and c) also provide some new insight. Consistent with the theoretical prediction that the CBM are located at the Q point, the $\kappa$ measurements show a dip at the CBM for both DL-MoSe$_2$ and DL-WSe$_2$[21-23]. Since the CBM in SL-WSe$_2$ is already located at the Q point, it is natural to expect that the CBM in DL-WSe$_2$ is also located at the Q point. But apparently, in DL-MoSe$_2$, the interlayer coupling is able to push the Q$_C$ point to lower energy but very close to K$_C$. This would imply a coupling strength of ~ 0.2 eV which is much smaller than the interlayer coupling at the $\Gamma$ point.

In table I we listed the majority and minority orbitals at these critical points to guide the discussion of general behavior of these critical points. As the interlayer coupling is mediated through the chalcogen atoms, its p-orbital orientation plays an important role. At Γ, the minority orbitals have $p_z$ character which explains the large $\Delta_{\Gamma-\Gamma}$. In contrast, the in-plane orientation of p-orbitals leads to little interlayer coupling for K states. At $Q_C$, the p-orbital is a mixture of $p_x$, $p_y$, and $p_z$, one thus expects some interlayer coupling but weaker than that at the Γ point, consistent with the experimental observation.

In summary, by using a comprehensive STS approach, we reveal the detailed electronic structures of transition metal dichalcogenides. In particular, we resolve the energy locations of different thresholds and their origins at different critical points in the BZ. We further show how these critical points are manifested when different transition metal atoms are used, from which we uncover that SL-WSe$_2$ has an indirect quasi-particle gap. The effects of interlayer coupling are also revealed. As the electronic structures of 2D materials can be strongly influenced by external perturbation (such as strain, dielectric environment, and carrier screening), the local probe with the capability to resolve critical point energies will have profound implications in advancing the scientific understanding of the TMDs as an emerging class of 2D electronic and optoelectronic materials.

## Methods

**Growth of 2D TMDs samples.** The preparation of WSe$_2$ crystal flakes by the vapour-phase reaction has been reported before[11]. In brief, high purity metal trioxides WO$_3$ was placed in a ceramic boat in the center of a furnace while graphite substrate was placed in the downstream side of the furnace, adjacent to the ceramic boat. Selenium powder was heated by a heating tape and carried by Ar or Ar/H$_2$ gas to the furnace heating center. The temperature of furnace was gradually raised from room temperature to the desired temperature, and cooled down naturally after the reaction had occurred. MoSe$_2$ was grown on freshly cleaved HOPG substrate using MBE in an ultra-high-vacuum (UHV) chamber which has a base pressure of ~ 1x10$^{-10}$ Torr. High purity Mo (99.95%) and Se (99.999%) were evaporated from a home-built e-beam evaporator and an effusion cell, respectively, with a ratio of 1:30. The graphite substrate was kept at 500 ℃, and the growth rate was about 0.3 layer/hour. The sample was annealed in a Se flux at 600 °C for 30 min after growth. Before STM studies, the CVD samples are cleaned in the UHV chamber (base pressure < 6 x $10^{-11}$ torr) by annealing the sample at 300°C for 6 hours. The MBE samples are transferred *in-situ* between the growth chamber and the STM chamber under UHV environment. See the Supplementary information for the details of preparation of MoSe$_2$ on graphene substrate.

**Scanning tunneling microscopy.** All STM investigations reported in this work were acquired at 77 K in a home-built UHV STM (base pressure < 6 x $10^{-11}$ torr) with the electrochemically etched W-tips. The $(\partial I/\partial V)_I$ spectra were taken at a constant tunneling current by using a lock-in amplifier with a modulation voltage of 10 mV and at a frequency of 925 Hz. In the meantime, the tip-to-sample distance $Z$ changes corresponding to the scanning of bias $V$ in order to keep the constant current. Thus, the $Z$-$V$ and $(\partial I/\partial V)_I$ vs. $V$ were acquired simultaneously. The regular $dI/dV$ was taken with the same lock-in setup but with the feedback off.

We adopted two methods to measure the tunneling decay constant $\kappa$ that equals to - $d$ ln$I/2dz$ ≡ - $(dI/dz)/2I_0$. First, the $dI/dZ$ can be acquired by using the lock-in amplifier (as reported in Ref. 25) with a $Z$-modulation amplitude of 0.02 nm and a modulation frequency of 925 Hz which is faster than the feedback time constant. This result in a small current modulation superimposed on the feedback set current $I_0$. If the sample bias is swept slowly, the $Z$-$V$ and $(\partial I/\partial Z)_I$ vs. $V$ can also be acquired at the same time. In addition, we can also measure the $I$-$Z$ spectroscopy at a certain sample bias ($Z$ was usually swept for 1 Å). Then the $\kappa$ can be deduced from the slop of ln$I$ vs. $Z$ curve. To get the $\kappa$ vs. bias plotting as shown in Fig 4c (left panel), this measurement was repeated at a series of sample biases.

**Acknowledgements**

This research was supported with grants from the Welch Foundation (F-1672), and the US National Science Foundation (DMR-1306878 and 1205275). L.J.L. thanks the support from Academia Sinica Taiwan, AOARD-134137 USA and KAUST, Saudi Arabia. C.K.S also thanks the National Science Council, Taiwan for financial supports for a visiting chair professorship at the National Tsing Hua University, Taiwan (NSC 102-2811-M-007-034). We also thank useful discussions with Professor Wang Yao of Hong Kong University.

**Author Contributions**

C.D.Z. carried out the STM measurement. C.K.S. advised on the experiment and provided input on the data analysis. C.K.S. and C.D.Z. wrote the paper with inputs from the other co-authors. Y.X.C prepared the MBE grown $MoSe_2$ sample. A. J. carried out the photoluminescence measurements. L.J.L. coordinated the CVD growth effort. M.Y.L. and J.K.H. performed the CVD growth of TMDs. P.C.M. and R.M.F. prepared the graphene on SiC substrates.

**Additional information**

**Competing financial interests**

The authors declare no competing financial interests.

**Figures captions**

**Figure 1 | STM images of TMDs grown on HOPG and the schematic drawing of the band diagram for SL-TMD materials. a**, **b** STM images of the single layer $MoSe_2$ and $WSe_2$, respectively. **a** 3.5 V, 7 pA, **b** 3.0 V, 7 pA. The inset in **a** shows the domain boundaries on SL-$MoSe_2$. The STS were carried out far away from the boundaries. The inset in **b** is the Moiré pattern seen on SL-$WSe_2$. **c.** The generic electronic structure of SL-TMDs. The VBM at K is split due to spin-orbit coupling are labeled as $K_2$ and $K_V$. A schematic of the first BZ is shown in the inset with the critical points labeled.

**Figure 2 | Tunneling spectroscopy of SL-$MoSe_2$ and the corresponding simulations**. **a.** $dI/dV$, **b**. $(\partial I/\partial V)_I$, **c**. $(\partial Z/\partial V)_I$ and **d**. $d\ln I/dz$ (decay constant $\kappa$). The left and right panels in **b-d** are corresponding to the valence and conduction bands. The states corresponding to key critical points in BZ ($\Gamma$, K, Q and $M^*$) are aligned with the dashed orange lines, and labeled on top of **a**. The M* refers to the local lowest point between $\Gamma$ and K, which is near M point. **e.** Simulations for the spectroscopy of valence band. The panels from top to bottom are the schematic four parabolic bands, simulated $d(\ln I)/\partial V$, $(\partial Z/\partial V)_I$ and kappa, respectively. The simulations for $MoSe_2$ on graphite are carried out with the following thresholds $\Gamma_V$ = -1.87 eV, $K_2$ = -1.72 eV and $K_V$ = -1.49 eV.

**Figure 3 | Tunneling spectroscopy of SL-$WSe_2$. a.** $dI/dV$, **b**. $(\partial I/\partial V)_I$, **c**. $(\partial Z/\partial V)_I$ and **d**. $d\ln I/dz$ (decay constant $\kappa$). The $K_2$ states can only be detected in **b-d**, and can hardly be seen in the regular $dI/dV$ measurements as shown in **a**. The sharp peak at conduction band edge in the right panel of **d** is due to the near degeneration of K and Q points. With the $\kappa$ measurements in the right panel of **d,** we confirm the indirect gap of SL-$WSe_2$ experimentally.

**Figure 4 | Tunneling spectroscopy for the conduction band of DL-TMDs.** The left and right columns in (a-c) correspond to the STS of DL-$MoSe_2$ and -$WSe_2$, respectively. **a.** $dI/dV$, **b.** $(\partial Z/\partial V)_I$, **c**. decay constant $\kappa$. Only the spectra of conduction bands are shown in **b** and **c**. For comparison, the $(\partial Z/\partial V)_I$ for single layers are displayed in **b** as well. The CBM are located at Q points for DL -$MoSe_2$ and -$WSe_2$.

**Table I | Orbital components of electronic states at critical points of Brillouin zone in SL-TMDs** (cited from Ref. 26 where M and X refer to the transition metal atom and chalcogen atom, respectively) **and the summary of energy differences between the critical points determined experimentally.**

**References:**

1. Mak, K. F., Lee, C., Hone, J., Shan, J. & Heinz, T. F. Atomically thin $MoS_2$: a new direct gap semiconductor. *Phys. Rev. Lett.* **105,** 136805 (2010).

2. He, K. L. *et al.* Tightly Bound Excitons in Monolayer $WSe_2$. *Phys. Rev. Lett.* **113**, 026803 (2014).

3. Ross, J. S. *et al.* Electrical control of neutral and charged excitons in a monolayer semiconductor. *Nat. Commun.* **4**, 1474,(2013).

4. Splendiani, A. *et al.* Emerging Photoluminescence in Monolayer $MoS_2$. *Nano Lett.* **10**, 1271-1275 (2010).

5. Yan, T. F., Qiao, X. F., Liu, X. N., Tan, P. H. & Zhang, X. H. Photoluminescence properties and exciton dynamics in monolayer $WSe_2$. *Appl. Phys. Lett.* **105**, 101901 (2014).

6. Zhang, Y. et al. Direct observation of the transition from indirect to direct bandgap in atomically thin epitaxial $MoSe_2$. *Nat. Nanotech.* **9**, 111-115 (2014).

7. Alidoust, N. et al. Observation of monolayer valence band spin-orbit effect and induced quantum well states in $MoX_2$. *Nat. Commun.* **5**, 4673 (2014).

8. Zhang, C. D., Johnson, A., Hsu, C. L., Li, L. J. & Shih, C. K. Direct imaging of the band profile in single layer $MoS_2$ on graphite: quasiparticle energy gap, metallic edge states and edge band bending. *Nano Lett.* **14**, 2443–2447 (2014)

9. Ugeda, M. M. et al. Giant bandgap renormalization and excitonic effects in a monolayer transition metal dichalcogenide semiconductor. *Nat. Mater.* **13**, 1091–1095 (2014).

10. Huang, Y. L. et al. Bandgap tunability at single-layer molybdenum disulphide grain boundaries. *Nat. Commun.* **6**, 6298 (2015).

11. Huang, J. K. et al. Large-area synthesis of highly crystalline $WSe_2$ monolayers and device applications. *ACS Nano* **8**, 923-930 (2014).

12. Tersoff, J. & Hamann, D. R. Theory and application for the scanning tunneling microscope. *Phys. Rev. Lett.* **50**, 1998-2001 (1983).

13. Tersoff, J. & Hamann, D. R. Theory of the scanning tunneling microscope. *Phys. Rev. B* **31**, 805-813 (1985).

14. Stroscio, J. A. & Kaiser, W. J. Scanning Tunnelling Microscopy. Ch. 4.5 (Academic Press, INC., San Diego, 1993).

15. Stroscio, J. A., Feenstra, R. M. & Fein, A. P. Electronic structure of the Si(111)2 × 1 surface by scanning tunneling microscopy. *Phys. Rev. Lett.* **57**, 2579-2582 (1986).

16. Ramasubramaniam, A. Large excitonic effects in monolayers of molybdenum and tungsten dichalcogenides. *Phys. Rev. B* **86**, 115409 (2012).

17. Zhu, Z. Y., Cheng, Y. C. & Schwingenschlogl, U. Giant spin-orbit-induced spin splitting in two-dimensional transition-metal dichalcogenide semiconductors. *Phys. Rev. B* **84**, 153402 (2011).

18. Xiao, D., Liu, G. B., Feng, W. X., Xu, X. D. & Yao, W. Coupled spin and valley physics in

monolayers of MoS$_2$ and Other Group-VI Dichalcogenides. *Phys. Rev. Lett.* **108**, 196802 (2012).

19. Zeng, H. L. et al. Optical signature of symmetry variations and spin-valley coupling in atomically thin tungsten dichalcogenides. *Nat. Sci. Rep.* **3**, 1608 (2013).

20. Sahin, H. et al. Anomalous Raman spectra and thickness-dependent electronic properties of WSe$_2$. *Phys. Rev. B* **87**, 165409 (2013).

21. Bradley, A. J. et al. Probing the role of interlayer coupling and coulomb interactions on electronic structure in Few-Layer MoSe$_2$ nanostructures. *Nano Lett.* **15**, 2594-2599 (2015).

22. Zhao, W. et al. Origin of indirect optical transitions in few-layer MoS$_2$, WS$_2$, and WSe$_2$. *Nano Lett.* **13**, 5627-5634 (2013).

23. He, J., Hummer, K. & Franchini, C. Stacking effects on the electronic and optical properties of bilayer transition metal dichalcogenides MoS$_2$, MoSe$_2$, WS$_2$ and WSe$_2$. *Phys. Rev. B* **89**, 075409 (2014).

24. Terrones, H. & Terrones, M. Bilayers of transition metal dichalcogenides: different stackings and heterostructures. *J. Mater. Res.* **29**, 373-382 (2013).

25. Kim, J. et al. Quantum size effects on the work function of metallic thin film nanostructures. *Proc. Natl. Acad. Sci. USA* **107**, 12761-12765 (2010).

26. Liu, G. B., Xiao, D., Yao, Y., Xu, X. & Yao, W. Electronic strutures and theoretical modelling of two-dimensional group-VIB transition metal dichalcogenides. *Chem. Soc. Rev.* **44**, 2643 (2015).

# Figure 1

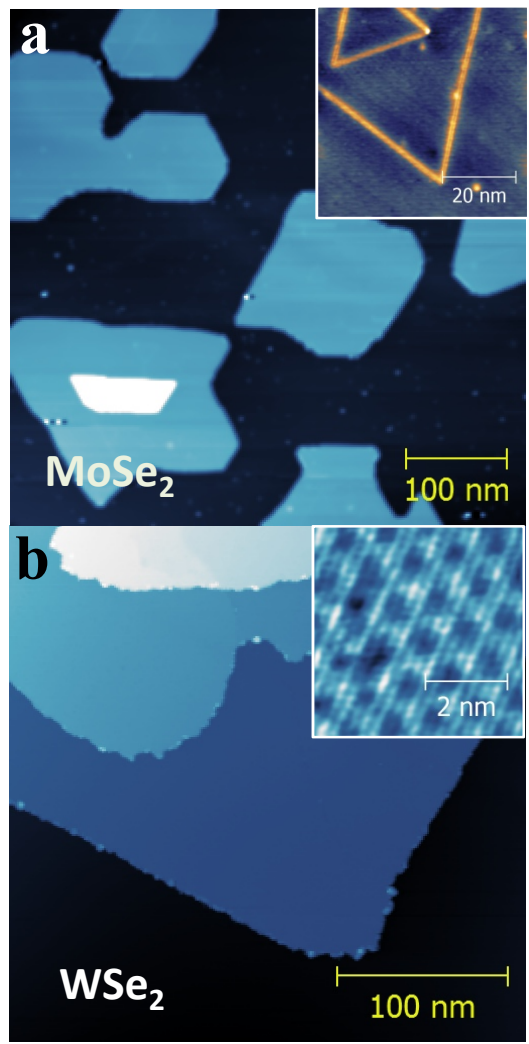
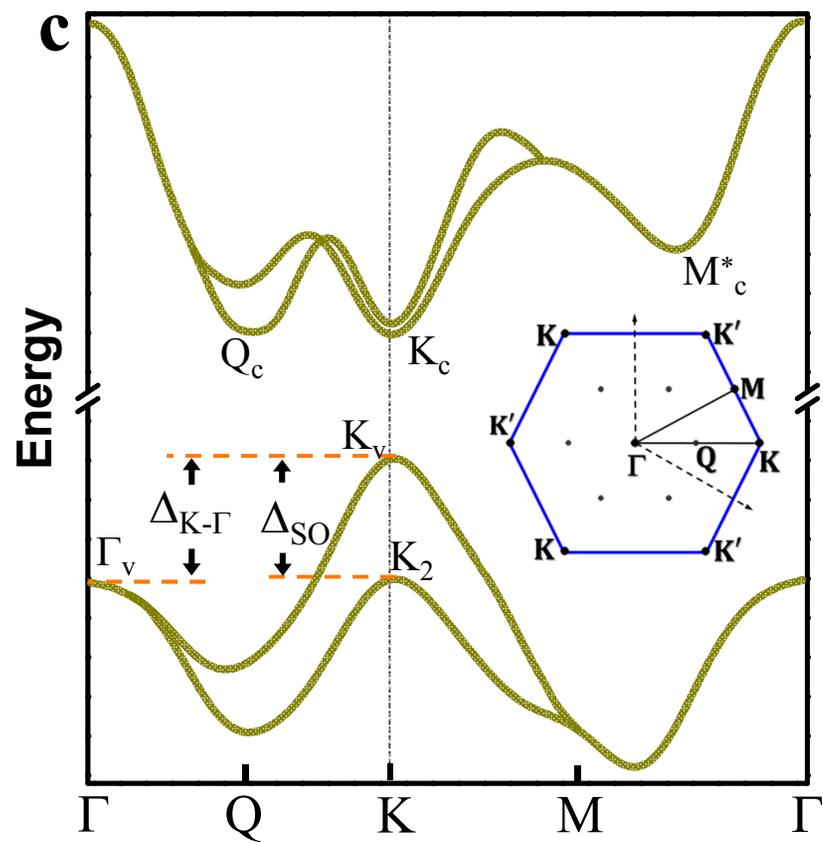

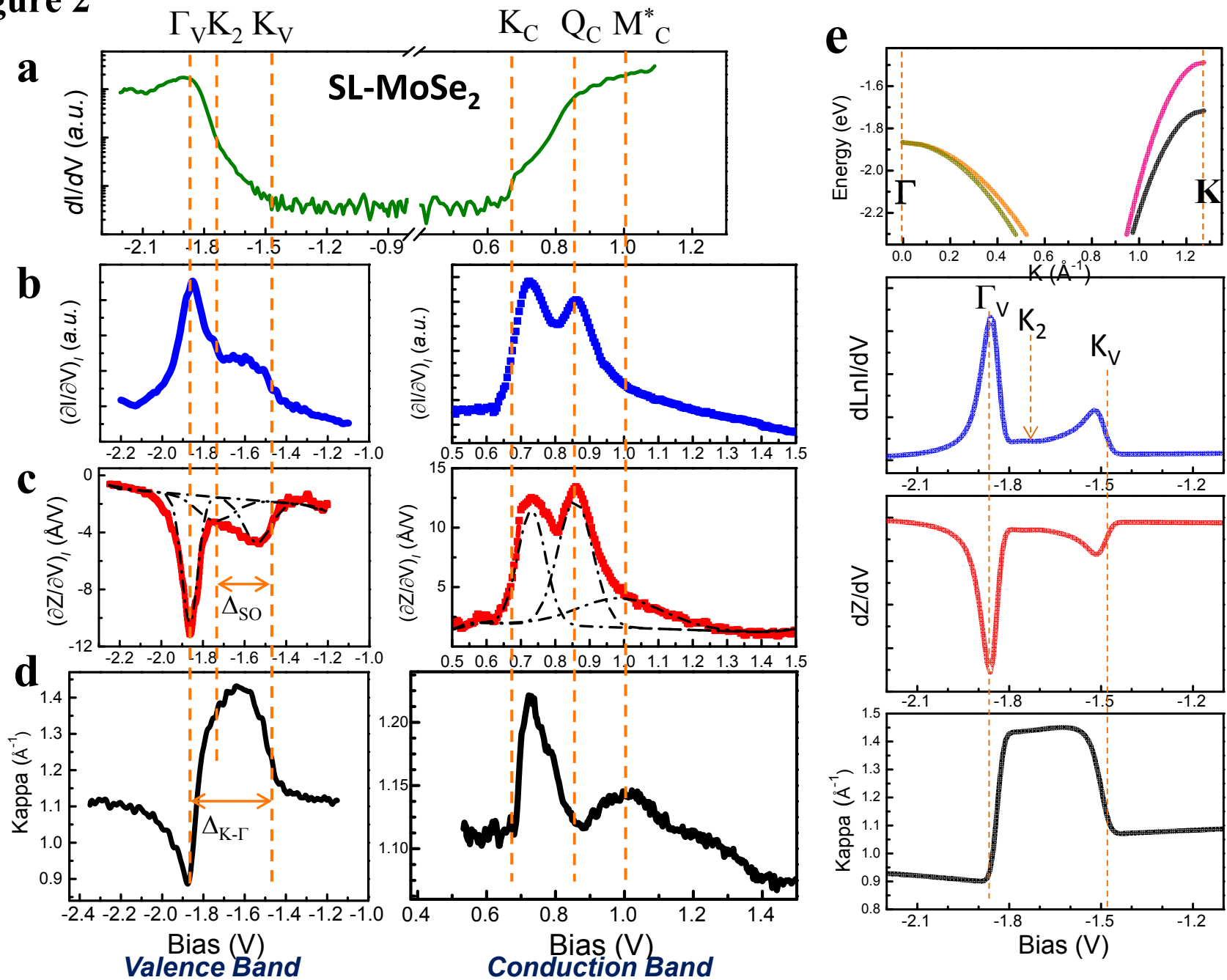

Figure 2

# Figure 3

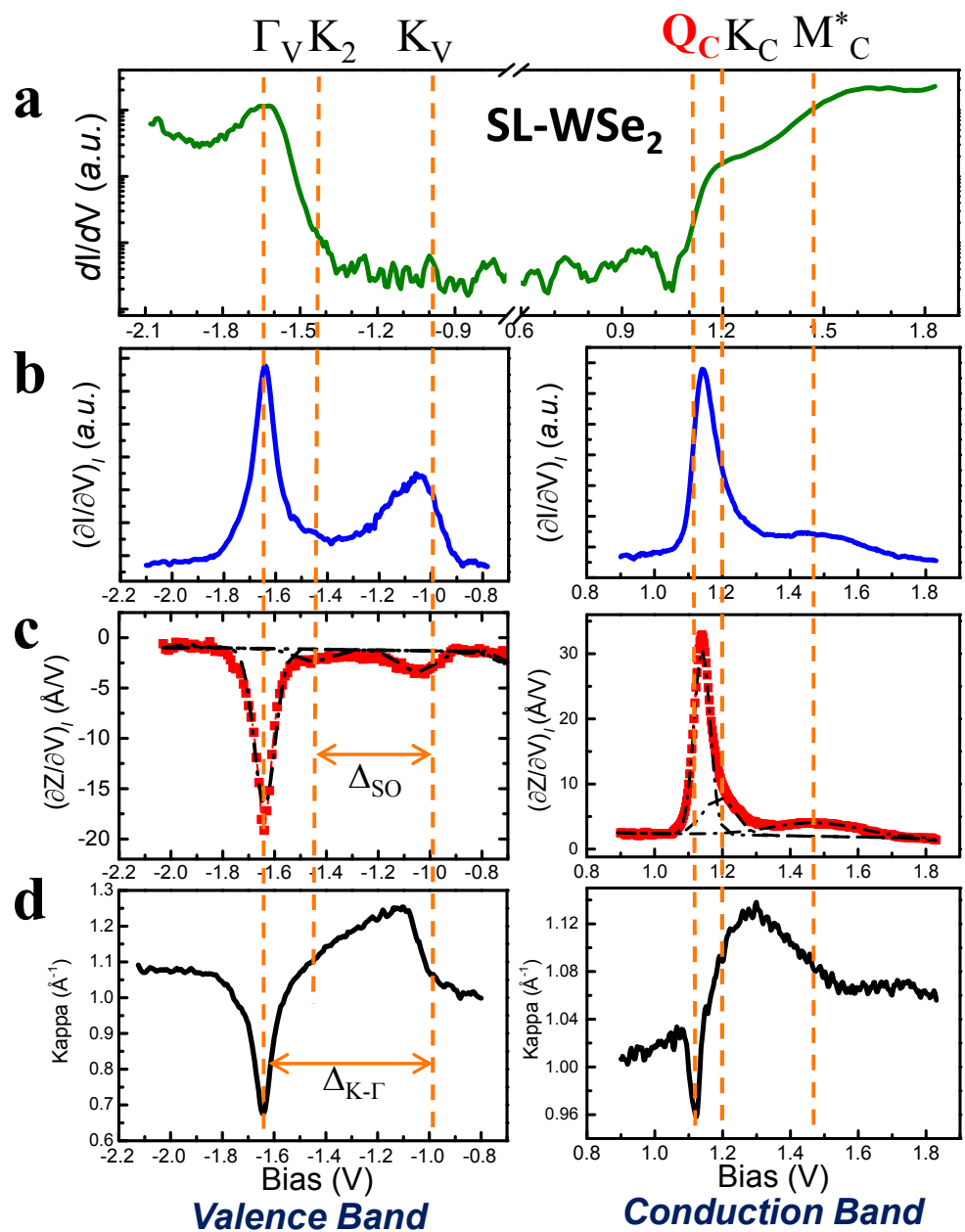

# Figure 4

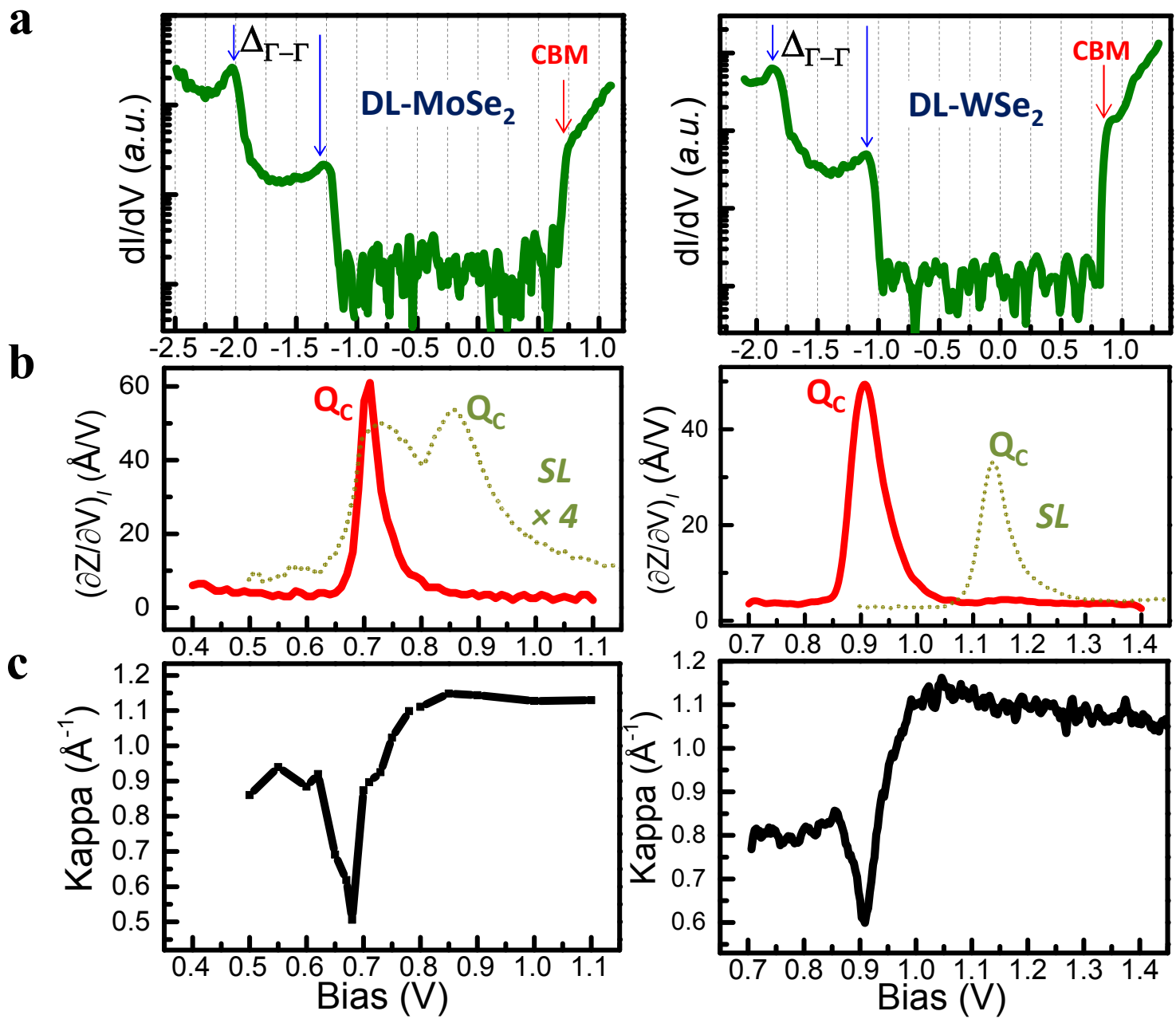

## Table I

| State | Majority of orbitals | Minority of orbitals |
|-------|----------------------|----------------------|
| $K_C$ | M-$d_{z^2}$ | X-$p_x, p_y$ |
| $Q_C$ | M–$d_{x^2-y^2}, d_{xy}$ | M-$d_{z^2}$, X-$p_x, p_y, p_z$ |
| $K_V$ | M–$d_{x^2-y^2}, d_{xy}$ | X-$p_x, p_y$ |
| $\Gamma_V$ | M-$d_{z^2}$ | X-$p_z$ |

| Energy difference (meV) | MoSe$_2$ | WSe$_2$ |
|---|---|---|
| $\Delta_{SO}$ | 240 | 440 |
| $\Delta_{K-\Gamma}$ (VB) | 390 | 640 |
| $\Delta_{Q-K}$ (CB) | 190 | -80 |

# Supplementary Information

## Numerical simulations

Simulations of experimentally measured quantities $(\partial Z/\partial V)_I$, $(\partial I/\partial V)_I$ and $\kappa = -\frac{1}{2}\frac{\partial \ln I}{\partial Z}$ are carried out using 4 parabolic bands for the TMD electronic states in the valence band. For $WSe_2$, as shown in Fig. S1a, the four bands are labeled as $\Gamma 1$, $\Gamma 2$, K1 and K2, with the effective mass of $m^*/m_o$ = 1.5, 1.3, 0.5, and 0.4, respectively. The thresholds used in the simulation for are $\Gamma_V$ = -1.65 eV, $K_2$ = -1.45 eV and $K_V$ = -1.05 eV. The results for $MoSe_2$ already appeared in the main text (Fig. 2e). The graphite DOS is modeled by a graphene-like DOS and is placed at 5Å below the TMD. The parabolic $E$ vs. $k$ dispersion results in a constant density of states proportional to the effective mass as shown in plot DOS for each). This approximation for the VB states of TMD greatly simplifies the calculations. Therefore, one can calculate the tunneling current by simply summing over contributions from individual bands.

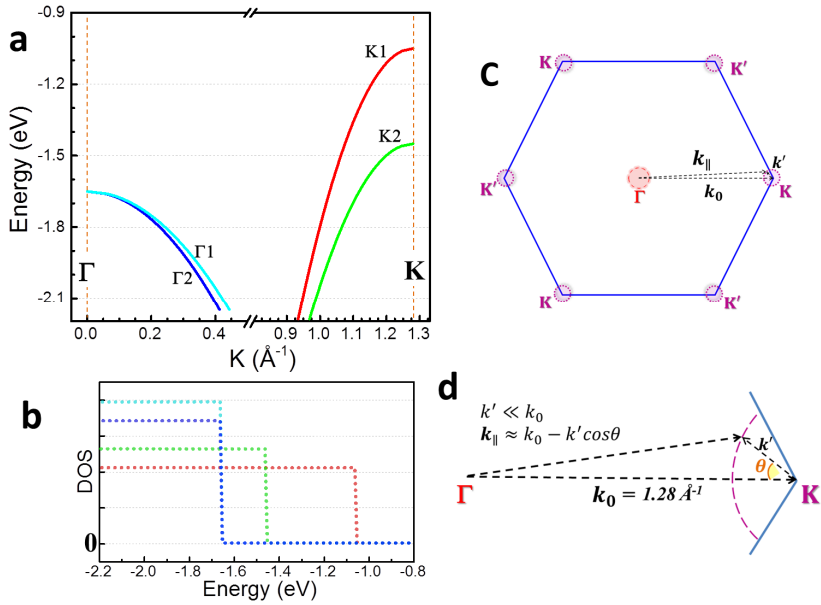

**Figure S1 a**. Four parabolic bands to simulate the spectroscopy for VB. And their corresponding DOS are shown in **b**. **c**. Schematic drawing of the geometry in BZ. **d** is a close-in plot.

The tunneling current can be written as

$$I_{total}(V) = I_{Gr}(V) + \sum_{i=1}^{4} I_i(V)$$

where $I_{Gr}(V)$ is the contribution from the underlying graphite evaluated at a distance 5 Å below that of TMDs and

$$I_i(V) = \int_{E_F}^{E_F+eV} \rho_i(E(k)) T(E(k), V) dE$$

is the contribution from each sub-bands used in the approximation. This expression assumes a constant DOS from the tip and ignores the finite width of the Fermi edge. The finite temperature effect of the Fermi Dirac distribution is taken into account by convoluting the density of the state with $-\partial f_{FD}/\partial E$.

The transmission function can be expressed as $T(E(k), V) = \exp(-2\tilde{\kappa} Z)$ where

$$\tilde{\kappa}(k_\parallel, V) = \sqrt{\frac{2m(\phi_o - \frac{e|V|}{2})}{\hbar^2} + k_\parallel^2}$$

Here the parallel momentum effect on the transmission function is taken into account explicitly in the expression for state-dependent decay constant $\tilde{\kappa}$. We use $\tilde{\kappa}$ to distinguish this quantity from the experimental observable of $-\frac{1}{2}\frac{\partial \ln I}{\partial Z}$. $\phi_o$ is taken to be 4 eV, a typical value used for an STM junction.

The parabolic $E$ vs. $k$ results in a circular constant energy surface (*i.e.* constant energy circle). For a constant energy circle surrounding the $\Gamma$ point, $k_\parallel$ is simply its radius. For the one surrounding the K point, we evaluate $k_\parallel$ by using the geometry as illustrated in Fig. S1b, d and obtain an expression for $k_\parallel = \sqrt{k_o^2 - 2k_o k' \cos\theta + k'^2}$ where $k_o$ is the crystal momentum from $\Gamma$ to K and $k'$ is the radius for the constant energy circle. We note however, specifically including the $\theta$-dependence in the evaluation, gives nearly identical result as if one just uses $k_\parallel = k_o - k'$.

To simulate $(\partial I/\partial V)_I$ one can directly use $\partial \ln I/\partial V$. Note that $\partial \ln I/\partial V = (\partial I/\partial V)/I$. Thus, at a constant current, $\partial \ln I/\partial V$ differs from $(\partial I/\partial V)_I$ only by a multiplication constant. Thus by carrying out the logarithmic derivative w.r.t. voltage numerically, we can simulate $(\partial I/\partial V)_I$ directly. This is shown in upper panels of Fig. 2e and Fig. S2a. The quantity $\kappa = -\frac{1}{2}\frac{\partial \ln I}{\partial Z}$ is obtained by simulating *I(V)* at two different Z values differing by 0.1 Å (in our case at $Z_1$ = 8 Å and $Z_2$ = 7.9 Å) to obtain $\frac{\partial lnI}{\partial Z}$. The resulting $\kappa = -\frac{1}{2}\frac{\partial lnI}{\partial Z}$ is shown in the lower panels of Fig. 2e and Fig. S2c. Since $(\partial Z/\partial V)_I = (\partial Z/\partial I)_I \cdot (\partial I/\partial V)_I$, we can extract this quantity directly from $(\partial lnI/\partial V)/(\partial lnI/\partial Z)$, the ratio of the two quantities already simulated.

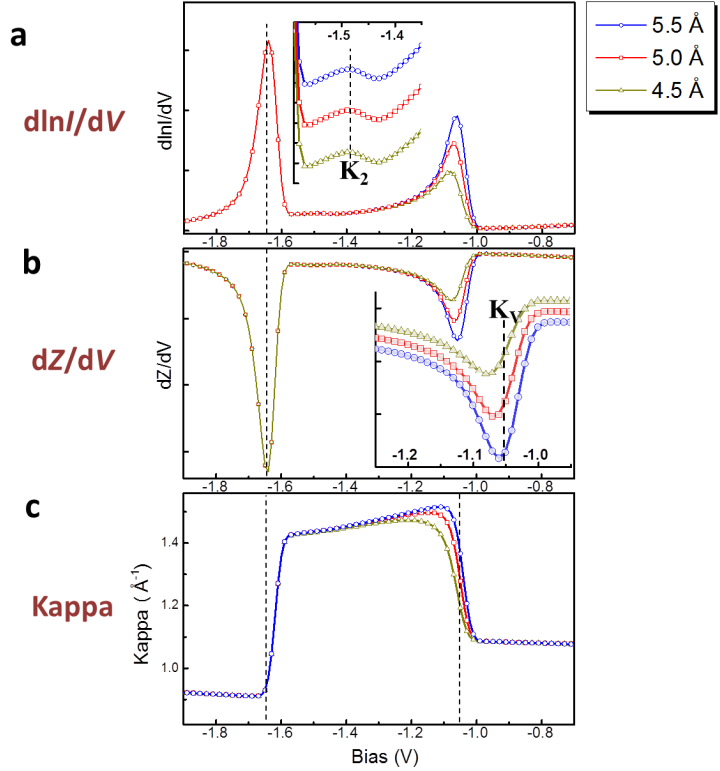

**Figure S2** Simulations for the valence band of SL-WSe$_2$ **a-c** are the simulated dln*I*/d*V*, $(\partial Z/\partial V)_I$ and kappa, respectively. The results for different distances to graphite are displayed in **a-c** with different color codes. The insets in **a** and **b** are the close-up plots for the $K_2$ and $K_V$, respectively.

The reason to choose graphite layer to be at 5 Å below is based on the fact that the DOS of TMD are concentrated in the TM layer (about 2 Å below the top Se layer) and the thickness of the TMD is about 7 Å. But the final result is rather insensitive to this choice of the physical location of graphite. The results for ΔZ = 4.5 Å and 5.5 Å are also included in the Fig. S2. As one can see, for $(\partial Z/\partial V)_I$ and $(\partial I/\partial V)_I$ the main difference is the amplitude of the peak (and dips). Moreover, the mid-point of the transition from TMD to graphite changes by about 0.01 eV when ΔZ is changed from 4.5 Å to 5.5 Å, much smaller than our experimental errors. The influence on the behavior of $\kappa = -\frac{1}{2}\frac{\partial \ln I}{\partial Z}$ is stronger. For this reason, we use primarily the $(\partial Z/\partial V)_I$ and $(\partial I/\partial V)_I$ curves to identify the location of $K_V$.

**Quasiparticle gap of SL-MoSe$_2$ on graphene**

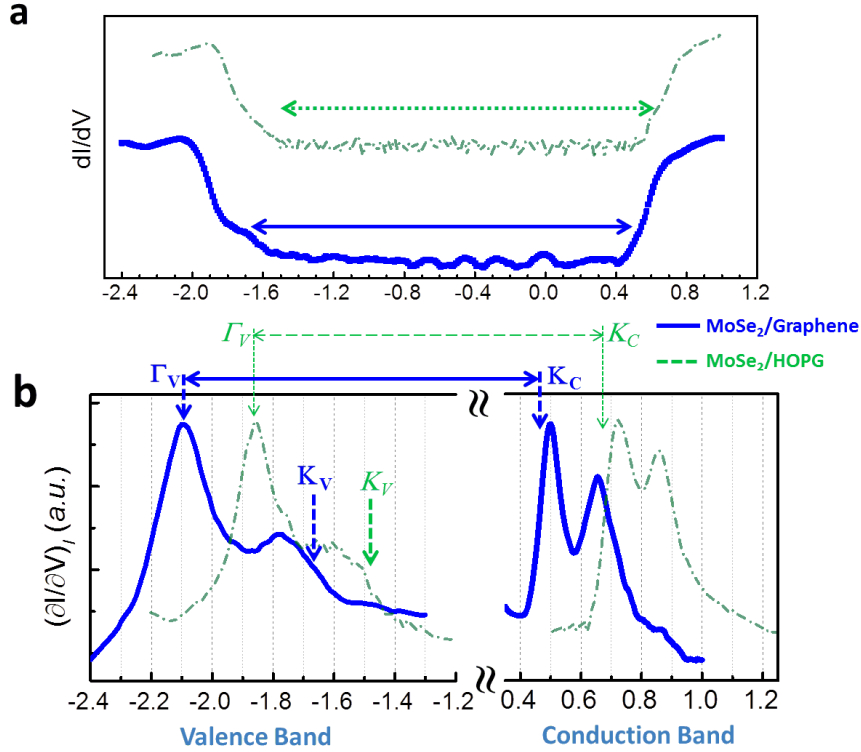

**Figure S3 a**. d$I$/d$V$ at constant $Z$ and **b**. $(\partial I/\partial V)_I$ of the SL-MoSe$_2$ grown on graphene. For comparison, the corresponding results on the HOPG substrate are displayed in green dashed lines.

The extreme two dimensional geometry of SL-TMDs makes them particularly interesting to study how the surrounding environment influences their electronic structures. Ugeda *et al.* has reported a significant reduction of the band gap of SL-MoSe$_2$ on graphite compared to SL-MoSe$_2$ on bilayer graphene [*Nat. Mater.* **13**, 1091 (2014)]. Here we investigate this issue using the refined method discussed in this manuscript.

The graphene was prepared on the Si-face of a SiC(0001) wafer by initially hydrogen etching that surface at 1620°C for 3 min to remove polishing damage. That procedure was followed by the graphene growth at 1590°C for 30 minutes in 1 atm argon environment [*J. Vac. Sci. Technol.*

*B* **28**, C5C1 (2010)]. Low-energy electron diffraction characterization revealed slightly less than 2 ML of graphene formed on the surface (i.e. in addition to the well-known carbon-rich "buffer layer" that forms between the graphene and the SiC). SL-MoSe$_2$ was then deposited on the graphene following the same procedure as discussed in the Method section.

In Figure S3, the results for SL-MoSe$_2$ grown on bi-layer graphene and on graphite are displayed in solid blue curves and dashed green curves respectively. The *dI/dV* spectra with constant Z are shown in Fig S3a. Compared with the HOPG substrate (green dashed), we could first see an obvious offset of both the VBM and CBM but they are shifted by equal amount. The constant current conductivity, $(\partial I/\partial V)_I$, as displayed in Fig S3b offers a much better quantitative comparison. The SL-MoSe$_2$ on graphene shows the identical features for both the valence band and conduction band. We assign the energy locations of $\Gamma_V$ point as 2.09 ± 0.03 eV, and K$_C$ as 0.46 ± 0.03 eV. Additionally, the K$_V$ can be assigned to be ~ -1.68 ± 0.05 eV. Compared to results of SL-MoSe$_2$ on HOPG, we observed only a rigid shift of the whole band structures (by about 0.22 eV), while the $\Delta_{\Gamma-K}$ (VB) and $\Delta_{Q-K}$ (CB) remained the same within the experimental error.

Thus based on our measurements of MBE SL-MoSe$_2$ on graphite and bi-layer graphene, the quasi-particle band gaps are identical with the same critical point energy separation. We do not dispute that the electronic properties of monolayer TMD films can influenced by its dielectric environment and the renormalization of quasi-particle band structures occurs. However we suggest that the similar coupling between SL-TMDs and the graphite/graphene substrates lead a similar renormalization effect at least in the samples we studied.

**Systematic trend of $\Delta_{Q-K}$ (CB)**

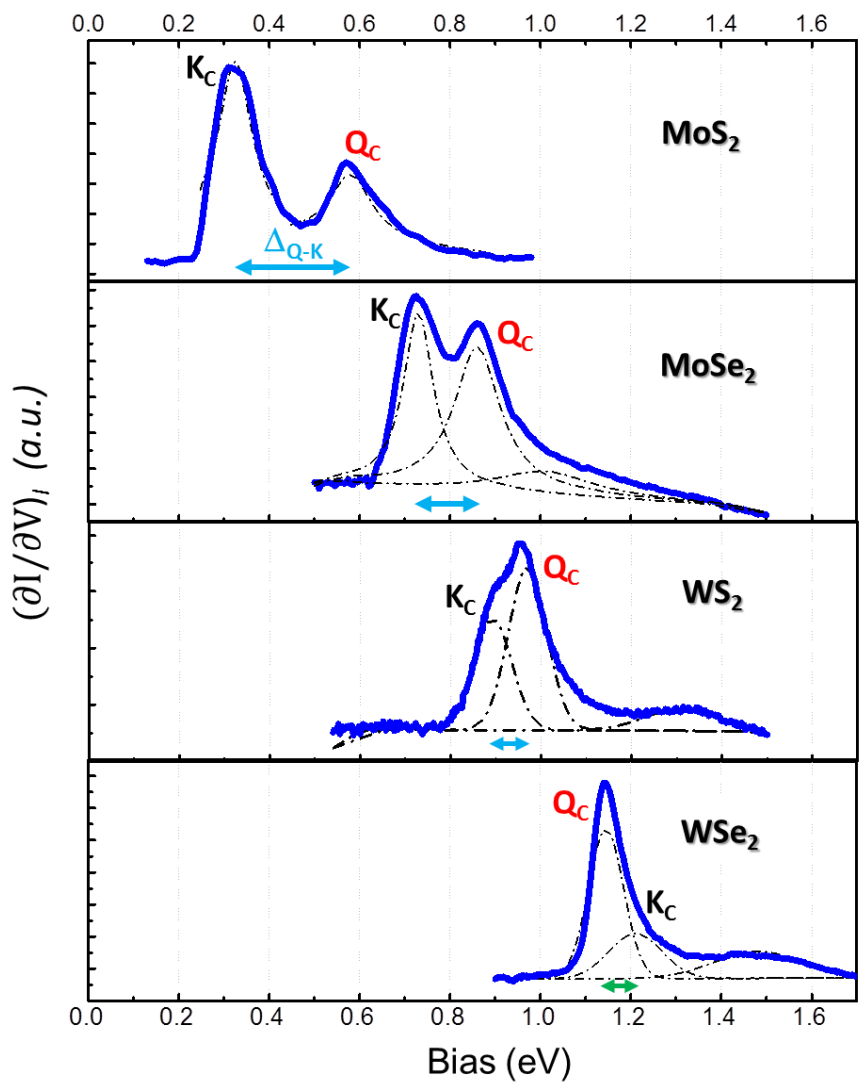

**Figure S4**

We carried out the similar STS investigations on two other TMDs compounds (*i.e.* the SL-$MoS_2$ and -$WS_2$), which are grown on HOPG substrates by CVD (see Methods section). The $(\partial Z/\partial V)_I$ spectra at conduction bands side are displayed in Fig. S4 for all four compounds. A systematic trend of the narrowing of $\Delta_{Q-K}$ is revealed, and in $WSe_2$ the energy levels of $Q_C$ and $K_C$ points get reversed eventually. This study further allows us to accurately assign the two thresholds reported earlier for conduction band structures in $MoS_2$ [*Nano Lett.* **14**, 2443–2447 (2014)]: The CBM is located at $K_C$ (~ 0.3 eV) while the second threshold (~ 0.55 eV) is assigned as $Q_C$.

**Figure S5 | Photoluminescence of MoSe$_2$ and WSe$_2$ on HOPG substrate**

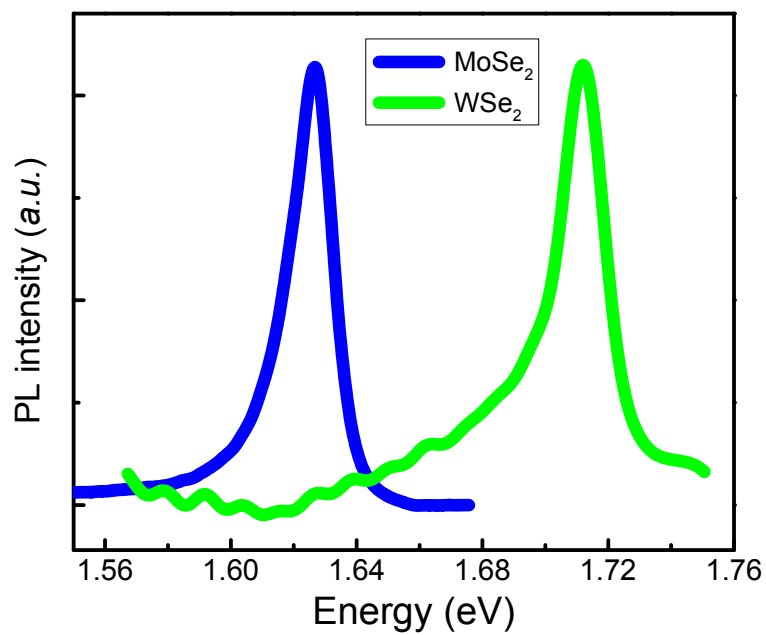

The measurements were carried out at 79 K. The 532 nm light was used to excite in glancing angle excitation geometry. The PL was collected with an optical microscope, and analyzed using an ARC Spectra Pro-500i spectrometer and a Si CCD detector. The transition energies of the neutral exciton A at 79 K are 1.63 eV and 1.71 eV for MoSe$_2$ and WSe$_2$, respectively.